\documentclass[superscriptaddress,twocolumn]{revtex4}

\usepackage{epsfig}

\begin{document}
\title{Dimensionality induced entanglement in macroscopic dimer systems}
\author{Dagomir Kaszlikowski}
\affiliation{Department of Physics, National University of
Singapore, 117542 Singapore, Singapore}

\author{Wonmin Son}
\affiliation{The School of Physics and Astronomy, University of
Leeds, Leeds, LS2 9JT, United Kingdom} \email{w.son@leeds.ac.uk}

\author{Vlatko Vedral}
\affiliation{Department of Physics, National University of
Singapore, 117542 Singapore, Singapore}
\affiliation{The School of
Physics and Astronomy, University of Leeds, Leeds, LS2 9JT, United
Kingdom}


\begin{abstract}
We investigate entanglement properties of mixtures of short-range
spin-s dimer coverings in lattices of arbitrary topology and
dimension. We show that in one spacial dimension nearest neighbour
entanglement exists for any spin $s$. Surprisingly, in higher
spatial dimensions there is a threshold value of spin $s$ below
which the nearest neighbour entanglement disappears. The traditional
``classical'' limit of large spin value corresponds to the highest
nearest neighbour entanglement that we quantify using the
negativity.
\end{abstract}
\maketitle

Consider a macroscopic system of $N$ subsystems with an arbitrary
spin $s$. Maximal entanglement of such a system corresponds to the
configuration where pairs of subsytems exist in maximally entangled
states (dimers).  The amount of entanglement quantified by the
relative entropy in this case is $\frac{N}{2}\log{(2s+1)}$. The
dimer configuration is also the most robust to noise in the sense
that one has to destroy entanglement between each pair of dimers in
order to destroy all entanglement. However, there are many dimers
configuration leading to the same maximal amount of entanglement.
Frequently in nature there is no reason why some of them should be
more likely than others, which is why they occur as superpositions
or mixtures of all possible dimer coverings.

Here we analyze how superposing and mixing of dimers affects the
robustness of their entanglement. We limit ourselves to short range
dimers because in practice forces generating entanglement between
subsystem are themselves specially short range (in momentum space
this, of course, need not be the case, a issue that is briefly
addressed at the end of the paper).

Recently, there has been a considerable interest in a possible link
between entanglement and properties of many-body systems
\cite{entanglement-condensed}. Although a role of entanglement in
description of such large macroscopic systems is not entirely clear
there are some promising results showing the connection between the
critical phenomena and entanglement \cite{entanglement-critical}.
Our present paper is a small contribution to this field.

Let us first consider a one dimensional lattice that is a union of
two sub-lattices $L_A$ and $L_B$. The sites belonging to the lattice
$L_A$ ($L_B$) are enumerated by odd (even) numbers. We assume that
the lattice is described by the so-called spin liquid, i.e., a state
without magnetic order
\begin{eqnarray}
&&|\tilde{\psi}\rangle = |(ab)_{12}\rangle|(ab)_{34}\rangle\dots|(ab)_{(2N-1)(2N)}\rangle\nonumber\\
&&+|(ab)_{(2N)1}\rangle|(ab)_{23}\rangle\dots|(ab)_{(2N-2)(2N-1)}\rangle\nonumber\\
&&=|c_1\rangle+|c_2\rangle,
\label{rvb}
\end{eqnarray}
with $|(ab)_{n(n+1)}\rangle =
\frac{1}{\sqrt{S}}\sum_{k=0}^{S-1}\gamma^{k b}_S|k\rangle_n
|k+a\rangle_{n+1}$, $S=2 s+1$, $a,b=0,1,\dots, S-1$,
$\gamma=\exp{(\frac{2i\pi}{S})}$ and the index $n$ refers to the
$n$-th site on the ring. The states $|(ab)_{n(n+1)}\rangle$ form a
generalized Bell basis for two particles with spin $s$, i.e., they
are complete and maximally entangled. In analogy with the spin
$\frac{1}{2}$ case a maximally entangled state between two sites is
called a {\it dimer} and the states $|c_1\rangle,|c_2\rangle$ are
called {\it dimer coverings}. We are interested in entanglement
properties of the state $|\tilde{\psi}\rangle$ in the thermodynamicl
limit, i.e., for  $N\rightarrow\infty$. Note that it makes no
difference in this limit whether we are superposing or mixing two
dimer coverings. As will be proven below all the phase information
is absent from the local entanglement properties, which we are
interested in.

We can prove the following facts for one dimensional case
\begin{enumerate}
\item There is always nearest neighbour entanglement for arbitrary value of $s$. Furthermore we present the negativity as the function of $s$ and show that
it increases with $s$.
\item The subset of even (or odd) sites does not contain entanglement.
\item The subset of even (or odd) sites is maximally entangled to the rest.
\end{enumerate}

{\it Proof of 1}. Let us first derive a density matrix for the
nearest neighbours. Due to the translational invariance of the state
it is enough to consider the density matrix $\rho_{12}$ of the first
two spins. We have
\begin{eqnarray}
&&  |c_1\rangle =\sum_{k_1,k_2,\dots,k_N}\frac{\gamma^{(k_1+k_2+\dots+k_N)b}}{\sqrt{S}^{N}}|k_,k_1+a\rangle_{12}\nonumber\\
&& |k_2,k_3,\dots,k_N\rangle_{357\dots}|k_2+a,k_3+a,\dots,k_N+a\rangle_{468\dots}\\
&&  |c_2\rangle =\sum_{k_1,k_2,\dots,k_N}\frac{\gamma^{(k_1+k_2+\dots+k_N)b}}{\sqrt{S}^{N}}|k_1+a,k_2\rangle_{12}\nonumber\\
&&
|k_2+a,k_3+a,\dots,k_N+a\rangle_{357\dots}|k_3,k_4,\dots,k_{N-1},k_1\rangle_{468\dots}\nonumber
\end{eqnarray}
After some tedious but straightforward algebra we get
\begin{eqnarray}
&&Tr_{345\dots(2N)}(|c_1\rangle\langle c_1|)=|(ab)\rangle\langle(ab)|\nonumber\\
&&Tr_{345\dots(2N)}(|c_2\rangle\langle c_2|)=\frac{1}{S^2}\\
&&Tr_{345\dots(2N)}(|c_1\rangle\langle
c_2|)=\frac{\gamma^{abN}}{S^N}|(ab)\rangle\langle
[(a+2Na)b]|.\nonumber
\end{eqnarray}

Thus, the un-normalized density matrix $\tilde{\rho}_{12}$ reads
\begin{eqnarray}
&&\tilde{\rho}_{12} = \frac{1}{S^2}+|(ab)\rangle\langle (ab)|\nonumber\\
&&+\frac{\gamma^{abN}}{S^N}|(ab)\rangle\langle [(a+2Na)b]|+h.c.
\end{eqnarray}
The trace of the matrix $\tilde{\rho}_{12}$ equals $M=2+2S^{1-N}
\cos{(\frac{2\pi abN}{S})}$ if the number of the sites is a
multiplicity of $S$, i.e., $2N=mS$ ($m$ is an integer) and $2$
otherwise.

In the thermodynamic limit the normalized state $\rho_{12}$ becomes
an equal mixture of the maximally entangled state $|(ab)\rangle$ and
the white noise
\begin{equation}
\rho_{12}^{N\rightarrow\infty}=\frac{1}{2S^2}+\frac{1}{2}|(ab)\rangle\langle
(ab)|.
 \end{equation}

In the similar way we can compute the density matrix between next
nearest neighbours, for instance, between the first and the third
spin

\begin{eqnarray}
&&\tilde{\rho}_{13} = \frac{2}{S^2} +\frac{1}{S^N}\\
&&~~ \times\sum_{k_1,k_2}\gamma^{-abN} |k_1,k_2\rangle\langle
k_2+2(N-1)a,k_1+2a|+h.c.,\nonumber
 \end{eqnarray}
which becomes the white noise in the thermodynamical limit because
the off diagonal elements rapidly vanish with $N$. It can be checked
that for larger separation one always gets the white noise for the
same reason, i.e., the off-diagonal elements vanish with $N$.

We finally calculate the negativity between nearest neighbour spins.
Since their state is a mixture of a maximally entangled state with
identity, the overall eigenvalues will be the same mixtures of the
eigenvalues of the identity and the maximally entangled state. This
is also true for the partially transposed state. It is therefore
easy to see that the negative eigenvalues of the partially
transposed state all have the absolute value of $(S-1)/2S^2$. The
number of negative eigenvalues of the partially transposed total
state is calculated to be $(S^2-S)/2$. The total negativity, which
measures the amount of entanglement in the state \cite{Vidal}, is
hence $S(S-1)^2/4S^2$. We see that for large spin value, the total
negativity grows linearly with the size of spin as claimed earlier.

{\it Proof of 2 and 3}. The density matrix $\rho_o(\rho_e)$
describing the subset of all the odd (even) sites has the following
form
\begin{eqnarray}
&&\tilde{\rho}_{o} = 2+\frac{\gamma^{-Nab}}{S^N}\sum_{k_1,k_2,\dots,k_N}|k_1,k_2,\dots,k_N\rangle\nonumber\\
&&\langle k_N+2a,k_1+2a,\dots,k_{N-1}+2a|+h.c.\\
&&\tilde{\rho}_{e} = 2+\frac{1}{S^N}\sum_{k_1,k_2\dots k_N}|k_1+a,k_2+a,\dots,k_N+a\rangle\nonumber\\
&&\langle k_2,k_3,\dots,k_N,k_1|+h.c.\nonumber
\end{eqnarray}
Both of them become the white noise in the thermodynamic limit,
which implies that there is no entanglement between any subset of
odd (even) sites. However, the set of all the odd sites is maximally
entangled with the set of all even sites.

The entanglement between the subsets of odd and even sites can be
seen already at the level of four sites. For instance, the state of
the first four sites is given by
\begin{eqnarray}
&&\tilde{\rho}_{1234} = |(ab)_{12},(ab)_{34}\rangle\langle (ab)_{12},(ab)_{34}|\nonumber\\
&&+\frac{1_1}{S}\otimes |(ab)_{23}\rangle\langle (ab)_{23}|\otimes\frac{1_4}{S}\\
&&+O(S^{2-N})|(ab)_{13},(ab)_{24}\rangle\langle
(ab)_{14},(ab)_{23}|+h.c.\nonumber
\end{eqnarray}
It is clear that the the sites $1$ and $3$ treated as one subsystem
are entangled to the subsystem consisting of the sites $2$ and $4$
but there is no genuine multipartite entanglement. The above formula
can be easily generalized for an arbitrary subsets containing even
and odd sites and it can be seen that as long as the size of the
subsets is fixed the subset is not genuinely multipartite entangled
in the thermodynamical limit.

There are situations in which different dimer coverings have the
same energy. This happens for example  in the Majumdar-Ghosh
Hamiltonian \cite{MG}, where we have nearest and next nearest
neighbour interactions. Unless there is some broken symmetry
mechanism each different covering will contribute to the overall
state with equal weight in the thermodynamic equilibrium. This means
that is more appropriate to consider mixture of dimer coverings
rather than their superpositions.

We now consider mixtures $\sigma$ of this type in any spatial
dimension

\begin{eqnarray}
&&\sigma = \frac{1}{M} \sum_{k=1}^{M} |c_k\rangle\langle c_k|,
\end{eqnarray}
where $ |c_k\rangle$ is $k$-th dimer covering and $M$ is the number
of all possible dimer coverings. For simplicity we present our
result in two dimensions. The generalization to higher dimensions is
straightforward.

We consider an infinite two dimensional square lattice that is an
union of two sub-lattices $L_A$ and $L_B$. A site belonging to the
sub-lattices $L_A$ have neighbours belonging to the sub-lattices
$L_B$ (coordination number is 4). The spin liquid state on the
lattice \cite{Long} is defined as the superposition of all possible
dimer coverings between the sub-lattices $L_A$ and $L_B$. The
question arises what the ratios with which we should mix the
coverings are.

\begin{figure}[h]
\begin{center}
\centerline{\includegraphics[width=3.4in]{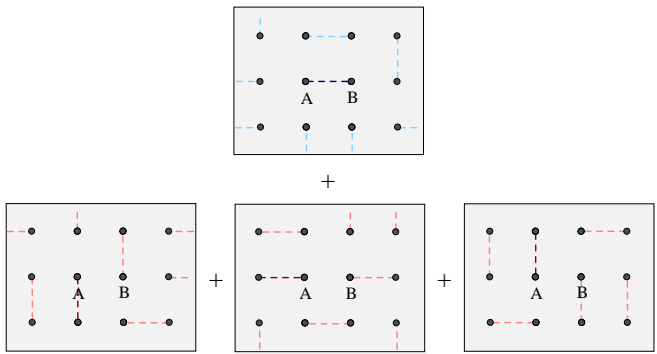}}
\end{center}
\caption{This figure illustrates the mixture of four possible dimer
coverings, $\{|c_k\rangle, k=1,\cdots, 4\}$, for an infinite lattice
site. The top box depicts one of the dimer coverings when the
neighbors A and B are maximally entangled. The other three show some
of the possible coverings when the state of A and B is maximally
mixed (therefore disentangled.) We argue that the ratio of the
number of the coverings when A and B is maximally entangled to the
number of coverings when A and B is maximally mixed is $1:3$ in the
thermodynamic limit (just like in the figure).} \label{fig1}
\end{figure}


To answer this question let us imagine a situation when two
arbitrary neighbours, say $i$ and $j$, are in a maximally entangled
state, i.e., they form a dimer. The rest of the sites can be covered
by dimers and there are $\Sigma$ of such coverings.  For an infinite
lattice $\Sigma$ is, of course, infinite but  we can assume that the
lattice is very large in which case $\Sigma$ is also large but
finite. Suppose now that the same neighbours are not in a maximally
entangled state, i.e., they form dimers with their other neighbours.
As is illustrated in Fig. (\ref{fig1}), it is easy to see that in
each such case (there are three of them) the number of  the
remaining dimer coverings $\Sigma'$ is equal and approximately the
same as in the previous case where the two neighbours $i$ and $j$
were in a maximally entangled state, i.e., $\Sigma\approx\Sigma'$.
Therefore, for an infinite lattice the ratio of the dimer coverings
including the sites $i$ and $j$ to the dimer coverings not including
the sites $i$ and $j$ equals $\frac{1}{3}$. The consequence of this
is that after tracing out all the other sites the density matrix
describing the neighbouring sites $i$ and $j$ ($i$ and $j$ can be,
in fact, any neighbours) is given by a Werner state containing
$\frac{3}{4}$ of the white noise. Thus, for a system consisting of
spin $\frac{1}{2}$ there is no entanglement between the neighbours.
The situation changes for larger spins $S>3$ (recall that $S=2s+1$,
where $s$ is the spin) because of the higher amount of the white
noise $\frac{S}{S+1}$ that can be admixed to a maximally entangled
state without destroying entanglement \cite{Werner-Horodecki}.

The similar reasoning can be applied to a ``honey-comb'' two
dimensional lattice (coordination number 3). In this case each spin
has only three neighbours which gives us the amount of the white
noise in the Werner state between the nearest neighbours equal to
$\frac{2}{3}$.  For spin $\frac{1}{2}$ we do not have entanglement
(the state is on the verge of being entangled) whereas entanglement
exists for larger spins, i.e, for $S>2$.

In higher than two dimensions we can apply exactly the same logic.
All we need to do is to calculate the ratio of the number of
coverings containing a dimer between two neighbouring sites and
those that do not. This ratio is always a fraction $\frac{1}{R}$,
where $R$ is a finite number that is the function of the
coordination number (for a simple regular lattice it equals to the
twice of the spatial dimension).  Therefore, in any dimension and
any lattice structure entanglement will always exist for spins with
magnitude higher than $R$; likewise it vanishes below this value.

A possible test of our prediction that there is a critical value of
the spin below which entanglement does not exist could go as
follows. Recently, condensates of fermionic atoms were observed in
the laboratory \cite{Jin}. In order to achieve condensation fermions
have to form entangled pairs (so called Cooper pairs), which then
behave like quasi-bosons. Entangled fermionic pairs are formed
through scattering process the strength of which can be controlled
experimentally by Feshbach resonance. Our calculation imply that
only atoms with sufficiently high spins can form entangled pairs. In
three dimensions we require atoms to have spin higher than
$\frac{5}{2}$. The experiment by Regal {\it et. al.} \cite{Jin} uses
potassium atoms, $^{40}K$, whose spin is $\frac{9}{2}$. Were they to
use atoms with spin of $\frac{5}{2}$ or lower, we conjecture that no
fermionic condensation would be possible.  Note that allowing long
range dimers can only increase the spin value necessary for
entanglement between any two sites on the lattice. This is because
the ratio of the dimer coverings contributing to maximally entangled
state between any two points on the lattice to the rest of the
coverings decreases.

\begin{figure}[t]
\begin{center}
\centerline{\includegraphics[width=2.4in]{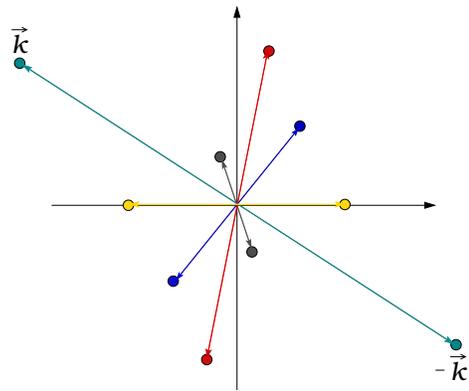}}
\end{center}
\caption{This figure depicts the arrangement of dimers in the BCS
ground state in the momentum space. The pairs of points represent
pairs of spin entangled electrons with momenta of the same magnitude
but of opposite directions. As we see this arrangement is not
isotrpoic, which is why our considerations no longer apply. In this
state we have fermionic pairs condensation and yet each of the
fermions is spin $\frac{1}{2}$. This is contrast to the isotrpoic
case when the minimal spin required for entanglement in three
dimensions is $\frac{7}{2}$ [See text for detailed explanation].}
\label{fig2}
\end{figure}


It is important to emphasize that our assumptions do not need to
hold in practice. For example, the BCS model of fermionic
condensation \cite{Bardeen57} has a ground state where momenta of
spins are coupled in opposite directions as is schematically shown
in Fig. (\ref{fig2}). Namely, spin dimers now exist only between
specific points on the lattice. This affects the ratio of mixing and
allows for entanglement with low spin in higher dimensions. This
model is highly non-isotropic in the sense of lattice points not
having the neighborhood.

Dimer coverings present very simple entangled states. Yet they are
physically relevant as they frequently occur in nature. Their
entanglement structure is simple. Here we show that overall
entanglement is fully characterized only in terms of nearest
neighbour entanglements. In spite of its simplicity, we encountered
one surprising fact. Namely, in higher than two dimensions the
existence of entanglement is dependent on the size of the spin in
each lattice site. Many questions remain open. For example, what
happens when a degree of anisotropy is introduced into dimer
coverings? Also, when does long range entanglement arise if we allow
long range dimers? We hope that our work stimulates a number of
other interesting directions of research.

{\it Acknowledgements.} We acknowledge M. Terra Cunha for useful
discussions related with this work. VV is grateful to the
Engineering and Physical Sciences Research Council in Uk and the
Royal Society and Wolfson Foundation for financial support.


\begin{thebibliography}{99}
\bibitem{Long}Long range spin $1/2$ dimer coverings, called spin gas, in two
dimensions and higher were considered in \cite{Chandran07}.
\end{thebibliography}
\end{document}